\documentclass[10pt,conference]{IEEEtran}
\usepackage{graphicx}   
\usepackage{adjustbox}  
\usepackage{float}      
\usepackage{caption}
\usepackage{amsmath}
\usepackage{amsfonts}
\usepackage{booktabs}

\makeatletter
\g@addto@macro\normalsize{%
  \setlength\abovedisplayskip{4pt plus 1pt minus 1pt}%
  \setlength\belowdisplayskip{4pt plus 1pt minus 1pt}%
  \setlength\abovedisplayshortskip{2pt plus 1pt minus 1pt}%
  \setlength\belowdisplayshortskip{2pt plus 1pt minus 1pt}%
  \setlength\jot{1.2pt}
}
\makeatother

\usepackage[font=footnotesize]{caption} 
\usepackage{subcaption}
\ifCLASSINFOpdf
\else
\fi
\begin{document}
%

\title{Channel Knowledge Map-Enabled NLoS ISAC Localization}
%
%
%

\author{
    \IEEEauthorblockN{
        Chentao Hong\textsuperscript{1}, 
        Di Wu\textsuperscript{1,2}, 
        Liang Wu\textsuperscript{1,2},
        Zaichen Zhang\textsuperscript{1,2} and Yong Zeng\textsuperscript{1,2}
    }

     \IEEEauthorblockA{\textsuperscript{1}National Mobile Communications Research Laboratory, Southeast University, Nanjing 210096, China\\}
    
    \IEEEauthorblockA{\textsuperscript{2}Purple Mountain Laboratories, Nanjing 211111, China}
    
    Emails: 
    \{220251213; studywudi; wuliang; zczhang; yong\_zeng\}@seu.edu.cn}

\maketitle

\begin{abstract}
Accurate localization in non-line-of-sight (NLoS) environments remains challenging even with both angle-of-arrival (AoA) and time-of-arrival (ToA) measurements. In complex urban scenarios, the absence of line-of-sight (LoS) paths and the lack of environment prior knowledge make geometric based localization methods inapplicable, while prior-based approach such as fingerprinting is sensitive to environmental perturbations. This paper proposes a novel environment-aware localization framework enabled by the emerging concept called channel knowledge map (CKM). In the offline stage, AoA–ToA path signatures are learned by the CKM, with each path mapped to one candidate scatterer, thereby forming geometric priors within the environment. In the online stage, observed paths are matched to the CKM to extract high-confidence scatterers. Nonlinear least squares (NLS) method is then applied to jointly estimate the user and dominant scatterer locations. Even with imperfect CSI matching, geometric feasibility consistent with CKM scatterer priors provides corrective information and suppresses ambiguity. Simulations demonstrate that the proposed scheme outperforms fingerprinting and offers a robust and scalable solution to address the challenging NLoS localization for integrated sensing and communication (ISAC) systems.

\end{abstract}


%
\IEEEpeerreviewmaketitle

\section{Introduction}
%
%
%
%
\thispagestyle{empty}

Integrated sensing and communication (ISAC) has been identified as a representative usage scenario within the IMT-2030 (6G) framework \cite{IMT2030}. By unifying radar-like sensing and data transmission on a shared waveform and infrastructure, ISAC enables environmental awareness and communication to coexist in a highly efficient manner \cite{isacTrend,ISACqianglong}. Its information-theoretic underpinnings \cite{ISACmutualInformation} and task-aware waveform \cite{ISACwaveform} are being actively studied. One of the key functionalities of ISAC is wireless localization, which plays a fundamental role in a wide range of applications, such as location-based services, emergency response coordination, transportation logistics, and indoor navigation \cite{xiao2022overview}. In addition, it is essential for supporting emerging 6G applications in the low-altitude economy, such as precise unmanned aerial vehicle (UAV) navigation and airspace coordination in urban environments \cite{song2024overviewcellularisaclowaltitude}.

Traditional wireless localization techniques typically assume the presence of a reliable line-of-sight (LoS) path between the base station (BS) and the user equipment (UE), and received signal strength (RSS), time of arrival (ToA), and angle of arrival (AoA) are usually employed \cite{RSSLocalization,ToAAoALocalization}. RSS-based methods are simple and cost-effective to implement, but their localization accuracy is limited due to susceptibility to shadowing and multipath effects. In contrast, ToA- and AoA-based methods exploit geometric constraints between the UE and multiple BSs and can achieve higher accuracy in multi-anchor deployments. However, in rich non-line-of-sight (NLoS) environments, the propagation paths become complex and time-varying, leading to significant degradation in the performance of geometry-based localization methods.

The challenges become more severe in the absence of angle of departure (AoD) measurements, as the UE’s transmit direction is typically uncontrolled and cannot be directly observed. Under these circumstances, conventional methods that rely on joint AoA–AoD geometric constraints in \cite{ruble2018wireless} become inapplicable. Although near-field processing can be beneficial by providing additional geometric constraints \cite{ZHIWEN}, theoretical studies in \cite{UnachievewithoutLos} show that under far-field NLoS conditions and without a priori channel knowledge, observations of only AoA and ToA remain insufficient. In such cases, the NLoS components make zero contribution to the equivalent Fisher information matrix (EFIM), rendering localization theoretically unachievable.

To address the challenge of achieving high-accuracy localization in NLoS environments, researchers have attempted to introduce prior information about scatterers to enhance localization performance. In \cite{old_scatter1} and \cite{old_scatter2}, the locations of the scatterers are assumed to follow a known probabilistic distribution, and UE localization is performed by leveraging this statistical model for optimal estimation. In addition, reconfigurable intelligent surfaces (RISs) have been proposed as a means to increase environmental controllability, thereby improving localization capability \cite{RISassist}. However, such approaches face practical limitation, as they depend on external infrastructure like RIS and it is often difficult to accurately model the spatial distribution of scatterers in real-world environments.

Some studies have attempted to adapt the concept of fingerprinting, originally developed for indoor environments, to outdoor localization. In this approach, a fingerprint database is constructed offline using signal features, and UE location is estimated online via similarity matching \cite{FPjournal}. Fingerprint-based localization encounters significant challenges in dynamic scenarios, which are subject to frequent dynamic changes due to factors such as traffic variations, and human activity \cite{fp3}, leading to unstable channel characteristic. This will reduce the discriminability of signal features and result in a single observation being indistinguishably associated with multiple reference points (RPs), thus causing ambiguity in localization and reducing localization robustness.

To address these limitations, in this paper, we propose an environment-aware localization framework based on the novel concept of channel knowledge map (CKM) \cite{ckm1}. CKM captures the intrinsic properties of multipath propagation in a site-specific manner and is originally proposed to enhance environmental awareness and to alleviate or even avoid the need for complex real-time CSI acquisition \cite{ckm2}. It has been applied in environment-aware NLoS sensing \cite{ckm_isac_nlos}, low-overhead channel estimation \cite{channelEstimation}, clutter suppression \cite{cluttersup}, UAV trajectory design \cite{UAVTrack} and RIS-aided communication \cite{RISCKM} and so on.

The core idea of the proposed framework is to treat multipath as a valuable source of geometric information. By systematically learning the angle and delay characteristics of multipath components across space, CKM embeds prior knowledge of the propagation environment. This approach achieves robust localization performance in complex NLoS environments while maintaining high physical interpretability and scalability, offering a new capability in NLoS scenarios. 
The principal contributions of this work are threefold. 
First, we exploit the environment awareness of a CKM to extract resolvable AoA-ToA pairs and infer scatterer priors. Second, we develop a similarity-based matching strategy that match the estimated multipath to the CKM to select high-confidence paths. Finally, we jointly estimate UE and dominant scatterer locations via nonlinear optimization that minimizes geometric-consistency and prior residuals. Simulation results are provided to show that the proposed method significantly outperforms fingerprint baselines across scenarios and maintains high accuracy under interference.

\section{System Model}

\begin{figure}[t]
    \centering
    \includegraphics[width=1\linewidth]{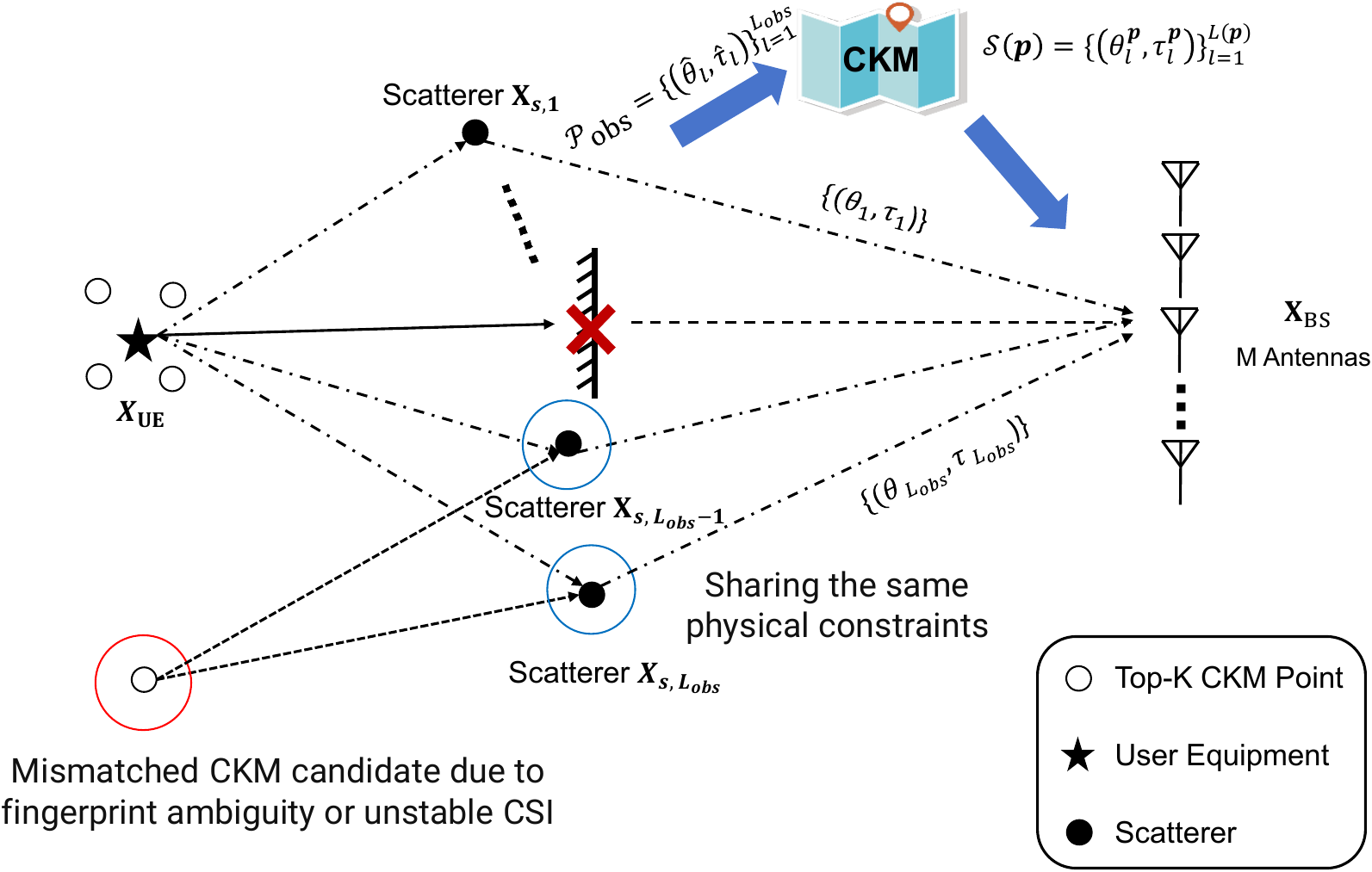}
    \captionof{figure}{An uplink SIMO ISAC system with CKM.}
    \label{fig:flowchart}
\end{figure}

As illustrated in Fig.~\ref{fig:flowchart}, we consider an uplink ISAC system where the UE is located at unknown location \( \mathbf{x}_{\mathrm{UE}} \in \mathbb{R}^2 \), and the BS, located at \( \mathbf{x}_{\mathrm{BS}} \in \mathbb{R}^2 \), is equipped with an \( M \)-element uniform linear array (ULA). The direct LoS path is blocked. Instead, the BS receives reflected signals from \( L \) dominant single-bounce scatterers located at \( \{ \mathbf{x}_{\mathrm{s},l} \in \mathbb{R}^2 \}_{l=1}^{L} \). For each path, the BS estimates the AoA $\theta_l$ and ToA $\tau_l$. Our goal is environment sensing and NLoS UE localization from $\{(\theta_l,\tau_l)\}_{l=1}^L$. 

To estimate the location of the UE, the following equations need to be solved
\begin{equation}
\label{eq:nlos_geom}
\begin{gathered}
\left\{
\begin{aligned}
&\lVert \mathbf{x}_{\mathrm{UE}}-\mathbf{x}_{\mathrm{s},l}\rVert_{2}
+\lVert \mathbf{x}_{\mathrm{s},l}-\mathbf{x}_{\mathrm{BS}}\rVert_{2}
= c\,\tau_l\\
&\mathbf{x}_{\mathrm{s},l}=\mathbf{x}_{\mathrm{BS}}+s_l\,\mathbf{u}(\theta_{l})
\end{aligned}
\right.
\\
\forall\, l=1,\ldots,L.
\end{gathered}
,
\end{equation}
where $\mathbf{u}(\theta_l) = [\cos\theta_l,\ \sin\theta_l]^T$ is the BS-anchored unit direction vector, and $s_l$ denotes the BS-to-scatterer radial distance along the AoA ray $\theta_l$.
In NLoS localization, the receiver estimates AoA/ToA pairs $\{(\theta_l,\tau_l)\}_{l=1}^L$ while the scatterer locations $\{\mathbf{x}_{\mathrm{s},l}\}_{l=1}^L$ and the UE location $\mathbf{x}_{\mathrm{UE}}$ are unknown, yielding an under-determined system with $2L+2$ unknowns but only $2L$ equations. This necessitates auxiliary information such as AoD or environmental priors. When UE has only one single antnena, AoD is unavailable. In this case prior-information–based methods become essential.

The dominant prior-based paradigms are fingerprinting and CKM. Fingerprinting performs a direct CSI match to a database. Let $\mathcal{D}=\{(\mathbf{x}_i,\mathbf{f}_i)\}_{i=1}^{N_{\text{ref}}}$ denote reference locations $\mathbf{x}_i$ with stored CSI vectors $\mathbf{f}_i$. Given an observation $\mathbf{f}$ at an unknown $\mathbf{x}_{\mathrm{UE}}$, a general fingerprinting estimator is
\begin{equation}
\label{eq:fp_general}
\widehat{\mathbf{x}}_{\mathrm{UE}}
= \arg\min_{\mathbf{x}_i\in\{\mathbf{x}_1,\dots,\mathbf{x}_{N_{\text{ref}}}\}}
d\!\big(\mathbf{f},\mathbf{f}_i\big),
\end{equation}
where $d(\mathbf{f},\mathbf{f}_i)$ is the Euclidean distance between CSI vectors.

In environments with strong multipath and layout symmetries, there may exist multiple distinct references $\mathbf{x}_i\neq\mathbf{x}_j$ such that
\begin{equation}
d(\mathbf{f},\mathbf{f}_i) \;\approx\; d(\mathbf{f},\mathbf{f}_j),
\end{equation}
which leads to fingerprint ambiguity as illustrated in Fig.~\ref{fig:flowchart}.

CKM augments CSI with environment-aware priors captured in AoA–ToA. These priors can resolve ambiguity by enforcing geometric consistency with the measured AoA-ToA pairs. 

The channel impulse response can be written as
\begin{equation}
\label{eq:compact_h}
\mathbf{h}(\tau)=
\sum_{l=1}^L\alpha_l\boldsymbol{\beta}(\theta_l)
\delta(\tau-\tau_l),
\end{equation}
where $\alpha_l$ is the complex channel gain, and the receive steering vector $\boldsymbol{\beta}(\theta_l)$ is defined for a ULA with the antenna array spacing  \( d \) as
\begin{equation}
\boldsymbol{\beta}(\theta_l) = 
\left[
1, \, e^{-j\frac{2\pi d}{\lambda} \sin\theta_l}, \, \dots, \, e^{-j\frac{2\pi d}{\lambda}(M-1)\sin\theta_l}
\right]^T,
\label{eq:steering_vector}
\end{equation}

The transmitted OFDM signal in the time domain is given by
\begin{equation}
s(t) = \sum_{\gamma=0}^{\Gamma-1} \sum_{n=0}^{N-1} b_{n,\gamma} e^{j2\pi n\Delta f \left(t-\gamma T_O - T_{\text{CP}}\right)} \text{rect} \left(\frac{t - \gamma T_O}{T_O}\right),
\end{equation}
where $\Gamma$ represents the total number of OFDM symbols, $N$ is the number of subcarriers, and $b_{n,\gamma}$ denotes the modulation symbol on the $n$-th subcarrier of the $\gamma$-th symbol. The symbol duration is $T_O = T + T_{\text{CP}}$, where $T$ is the effective symbol duration and $T_{\text{CP}}$ is the cyclic prefix (CP) duration. The parameter $\Delta f$ represents the subcarrier spacing, and $\text{rect}(\cdot)$ is the rectangular pulse-shaping function.

At the BS, the received signal is a superposition of multipath components and can be expressed as
\begin{equation}
\mathbf{y}(t)=\sum_{l=1}^L\alpha_l\boldsymbol{\beta}(\theta_l)s(t-\tau_l)+\mathbf{z}(t),
\label{eq:received_vector}
\end{equation}
where \( \mathbf{z}(t)\) is the additive Gaussian noise with power \( \sigma^2 \).


\section{NLoS Localization Enabled by CKM}
\subsection{Basic Principle of CKM}
The CKM serves as a spatial database that stores the angle–delay characteristics of the multipath environment. Specifically, it defines a mapping from each known transmitter location to its corresponding set of AoA–ToA pairs. Let \(\mathbb{P}\subset\mathbb{R}^2\) denote the set of potential UE locations \(\mathbf p\). The CKM is expressed as
\begin{equation}
\mathcal{M}_{\text{CKM}} : \mathbf{p} \in \mathbb{P} \;\mapsto\; \mathcal{S}(\mathbf{p}),
\label{eq:ckm_mapping}
\end{equation}
where
\begin{equation}
\mathcal{S}(\mathbf{p}) = 
\left\{ 
  \left( \theta_l^{\mathbf{p}}, \tau_l^{\mathbf{p}} \right) 
\right\}_{l = 1}^{L(\mathbf{p})},
\label{eq:ckm_entry}
\end{equation}
and \( L(\mathbf{p}) \) is the number of resolvable multipath components at location \( \mathbf{p} \).

When constructing the CKM, the most common approach is to directly traverse all UE locations within the region of interest. Accordingly, we partition the target area into a grid and sample reference UE locations to collect uplink signals, so that each location obtains representative angle--delay information. We follow the process in \cite{ISACqianglong} to obtain AoA-ToA pairs.

Although the CKM itself does not store scatterer coordinates, each entry at location $\mathbf p$ can be mapped, via a closed-form
geometric relationship, to a set of equivalent single-bounce scatterers. Specifically, at grid node $\mathbf p$, the UE is placed at
$\mathbf{x}_{\mathrm{UE}}=\mathbf p$ and the BS location $\mathbf x_{\mathrm{BS}}$ is known.
For the $l$-th CKM entry $\big(\theta_l^{\mathbf p},\tau_l^{\mathbf p}\big)$, define
$\mathbf r=\mathbf{x}_{\mathrm{UE}}-\mathbf x_{\mathrm{BS}}$ and the BS ray direction
$\mathbf u(\theta_l^{\mathbf p})=[\cos\theta_l^{\mathbf p},\,\sin\theta_l^{\mathbf p}]^{T}$.
Under a single-bounce model in~\eqref{eq:nlos_geom}, the scatterer can be calculated as
\begin{equation}
\;
{\mathbf x}_{\mathrm s,l}^{\mathbf p}
= \mathbf x_{\mathrm{BS}}
+ \frac{(c\,\tau_l^{\mathbf p})^2 - \|\mathbf r\|_2^{2}}
       {2\big(c\,\tau_l^{\mathbf p} - \mathbf r^{ T}\mathbf u(\theta_l^{\mathbf p})\big)}
  \; \mathbf u(\theta_l^{\mathbf p}).
\;
\label{eq:ckm_scatterer_single_formula}
\end{equation}

In the following subsections, these CKM–implied scatterers ${\mathbf x}_{\mathrm s,l}^{\mathbf p}$ act as environment priors that we fuse with real-time AoA/ToA in a coherent pipeline.

\subsection{Coarse Localization by Angle–Delay Similarity}
During online processing, the BS extracts the currently observed $L_{obs}$ resolvable multipath components as
\begin{equation}
    \mathcal{P}_{\mathrm{obs}}=\big\{(\hat{\theta}_l,\hat{\tau}_l)\big\}_{l=1}^{L_{obs}},
    \label{eq:obs_set}
\end{equation}
where $\hat{\theta}_l$ and $\hat{\tau}_l$ denote the estimated AoA and ToA of the $l$-th observed path.

Define the receive steering vector \(\boldsymbol{\beta}(\theta)\!\in\!\mathbb{C}^{M\times 1}\) as in the system model, and introduce the per–subcarrier delay steering vector
\begin{equation}
\mathbf{v}(\tau)=\left[1,e^{-j2\pi\Delta f\tau},\ldots,e^{-j2\pi(N-1)\Delta f\tau}\right]^T\in\mathbb{C}^{N\times1},
\end{equation}
which captures the deterministic phase progression across subcarriers induced by a delay \(\tau\).
We then form the virtual angle–delay snapshots for the observation and for a CKM grid \( \mathbf{p}\), which can be expressed as
\begin{equation}
\begin{cases}
\tilde{\mathbf{X}}_{\mathrm{obs}}
= \sum\limits_{l=1}^{L_{obs}} \boldsymbol{\beta}\!\left(\widehat{\theta}_l\right)\mathbf{v}(\hat{\tau}_l)^{H} \\
\tilde{\mathbf{X}}_{\mathrm{ckm(\mathbf{p})}}
= \sum\limits_{l=1}^{L(\mathbf{p})} \boldsymbol{\beta}\!\left(\theta_l^{\mathbf{p}}\right)\mathbf{v}(\tau_l^{\mathbf{p}})^{H}
\end{cases}
.
\end{equation}

Let $\mathbf{W}_{\!\theta}\!\in\!\mathbb{C}^{M\times N_\theta}$ and $\mathbf{V}_{\!\tau}\!\in\!\mathbb{C}^{N\times N_\tau}$ denote the spatial and delay DFT dictionaries, where $M$ and $N$ are the numbers of array elements and subcarriers, and $N_\theta$ and $N_\tau$ are the numbers of DFT grid points along the angular and delay axes. The 2-D spectrum $\mathbf{B}_x$ and power map $\mathbf{P}_x$ are defined as
\begin{equation}
\begin{cases}
\mathbf{B}_x = \mathbf{W}_{\!\theta}^{H}\,\tilde{\mathbf{X}}_{x}\,\mathbf{V}_{\!\tau}\\
\mathbf{P}_x = \mathbf{B}_x \odot \mathbf{B}_x^{*}
\end{cases}.
\label{eq:Bx_def}
\end{equation}
Here the subscript $x\in\{\mathrm{obs},\,\mathrm{ckm}(\mathbf{p})\}$ denote either the observation or the CKM corresponding to a candidate location $\mathbf{p}$, the notation $\odot$ denotes the Hadamard product.
In the joint analysis over the array domain and the delay domain, the spatial and delay DFT dictionaries are employed as transformation bases.  
The spatial dictionary $\mathbf{W}_{\!\theta}$ projects the measurements collected over $M$ array elements onto $N_\theta$ discrete angular grid points, while the delay dictionary $\mathbf{V}_{\!\tau}$ projects the $N$ frequency-domain samples over subcarriers onto $N_\tau$ discrete delay grid points.  
For any data block $\tilde{\mathbf{X}}_{x}$, performing a two-sided projection onto these dictionaries yields its two-dimensional angle–delay spectrum $\mathbf{B}_x$ and the corresponding power map $\mathbf{P}_x$.

\begin{figure}[t]
\centering
 \includegraphics[width=0.9\linewidth]{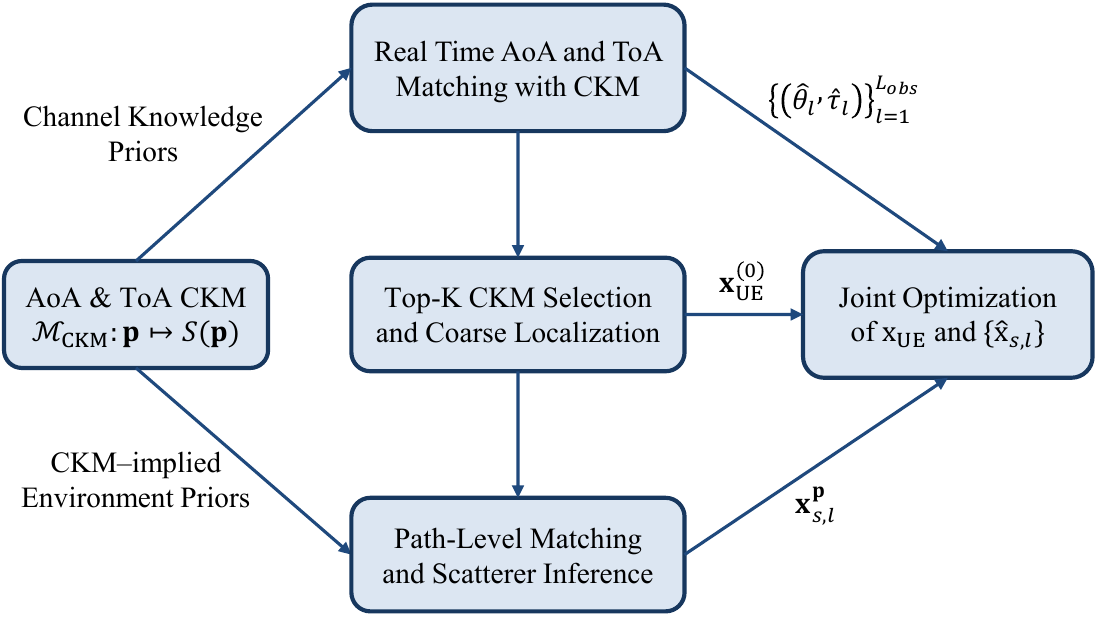}
    \captionof{figure}{CKM-enabled localization framework.}
    \label{fig:Total_flowchart}
\end{figure}

To align continuous physical parameters with the discrete DFT grid, we define the angular and delay phase offsets as
$\varphi_{\theta,l}^{(x)} = 2\pi\!\big(\tfrac{d}{\lambda}\sin\theta_{l}^{(x)}-\tfrac{\kappa}{N_\theta}\big)$ and
$\psi_{\tau,l}^{(x)} = 2\pi\!\big(\Delta f\,\tau_{l}^{(x)}-\tfrac{n}{N_\tau}\big)$.
Given $x\in\{\mathrm{obs},\,\mathrm{ckm}(\mathbf{p})\}$, the 2-D spectrum at bin $(\kappa,n)$ is
\begin{equation}
\begin{aligned}
\big[\mathbf{B}_x\big]_{\kappa,n}
&= \frac{1}{\sqrt{MN}} \sum_{l=1}^{L_x}
e^{-j\,\frac{M-1}{2}\,\varphi_{\theta,l}^{(x)}}
\frac{\sin\!\big(\tfrac{M}{2}\,\varphi_{\theta,l}^{(x)}\big)}
     {\sin\!\big(\tfrac{1}{2}\,\varphi_{\theta,l}^{(x)}\big)}
\\[-2pt]
&\qquad\qquad\times\;
e^{-j\,\frac{N-1}{2}\,\psi_{\tau,l}^{(x)}}
\frac{\sin\!\big(\tfrac{N}{2}\,\psi_{\tau,l}^{(x)}\big)}
     {\sin\!\big(\tfrac{1}{2}\,\psi_{\tau,l}^{(x)}\big)} \, ,
\end{aligned}
\end{equation}
and the corresponding power map is
$\big[\mathbf{P}_x\big]_{\kappa,n} = \big|\big[\mathbf{B}_x\big]_{\kappa,n}\big|^{2}$.
This unified formulation places both the observation and the candidate CKM at location $\mathbf{p}$ on the same angle–delay grid with consistent normalization, enabling direct comparison and robust matching. Subsequently, peak-normalized maps can be defined as
\(P_x^{\prime}=\frac{P_x}{\max (P_x)}\).
The similarity at grid $\mathbf{p}$ is
\begin{equation}
\mathrm{Sim}(\mathbf{p})=\frac{\mathrm{vec}\left(P_{\mathrm{obs}}^{\prime}\right)^{T}\mathrm{vec}\left(P_{\mathrm{ckm(\mathbf{p})}}^{\prime}\right)}{\|\mathrm{vec}(P_{\mathrm{obs}}^{\prime})\|_{2}\|\mathrm{vec}(P_{\mathrm{ckm}(\mathbf{p})}^{\prime}\|_{2}}.
\label{eq:sim_color_cos}
\end{equation}

We first form a global CKM candidate set by scanning the entire grid domain
\(\mathbb{P}\subset\mathbb{R}^2\) of angle–delay indices. Let \(K_{\mathrm{cand}}\in\mathbb{N}\) be the number of candidates to keep.
We rank all grids by the cosine similarity \(\mathrm{Sim}(\mathbf{p})\) in (\ref{eq:sim_color_cos}) and
retain the top-\(K_{\mathrm{cand}}\) elements
\begin{equation}
\mathcal{G}_{\mathrm{cand}}
=\operatorname{Top}_{K_{\mathrm{cand}}}\Big\{\,\mathrm{Sim}(\mathbf{p})\ \Big|\ \mathbf{p}\in\mathbb{P}\,\Big\}.
\label{eq:cand_topk_global}
\end{equation}

We then initialize the UE location by the similarity-weighted barycenter of these
candidates
\begin{equation}
\mathbf{x}_{\mathrm{UE}}^{(0)}=
\frac{\displaystyle\sum_{\mathbf{p}\in\mathcal{G}_{\mathrm{cand}}}\mathrm{Sim}(\mathbf{p})\,\mathbf{p}}
{\displaystyle\sum_{\mathbf{p}\in\mathcal{G}_{\mathrm{cand}}}\mathrm{Sim}(\mathbf{p})}.
\label{eq:init_bary_global}
\end{equation}

This step serves as a WKNN-style fingerprinting initializer, where the fingerprint is the AoA–ToA pattern provided by the CKM. It narrows the search region and improves convergence for the nonconvex objective in in~\ref{estimation}, reducing the chance of spurious minima under complex scenario.

\subsection{Path-Level Matching and Weighted Prior Selection}
In the previous subsection, we evaluated grid-wise similarity on the angle–delay power maps and obtained $\mathcal{G}_{\mathrm{cand}}$. 
To turn spectral evidence into path-level cues, we pair the observed path
$\{(\hat{\theta}_l,\hat{\tau}_l)\}_{l=1}^{L_{\mathrm{obs}}}$ with CKM candidate paths
$\{(\theta_m^{\mathbf{p}},\tau_m^{\mathbf{p}})\}_{m=1}^{L(\mathbf{p})}$.
Angles and delays are mapped directly onto the angle–delay power map via
$\kappa(\theta)=\tfrac{N_\theta d}{\lambda}\sin\theta$ and
$\pi(\tau)=N_\tau\Delta f\,\tau$,
where $\kappa(\cdot)$ and $\pi(\cdot)$ denote respectively the continuous coordinates on the map’s angle and delay axes.
Given an observed path $(\hat\theta_l,\hat\tau_l)\in\mathcal{P}_{\mathrm{obs}}$ and a CKM path
$(\theta^{\mathbf{p}}_m,\tau^{\mathbf{p}}_m)\in\mathcal{S}(\mathbf{p})$, we define the nonnegative dissimilarity
\begin{equation}
D_{l,m}(\mathbf{p})
=\big\|[\kappa(\hat\theta_l),\,\pi(\hat\tau_l)]^{T}
      -[\kappa(\theta^{\mathbf{p}}_m),\,\pi(\tau^{\mathbf{p}}_m)]^{T}\big\|_2 ,
\label{eq:D_index}
\end{equation}
where a smaller $D_{l,m}(\mathbf{p})$ indicates a better match. 

For any fixed $\mathbf{p}\in\mathcal{G}_{\mathrm{cand}}$, we perform a set-wise one-to-one assignment between all observed paths and CKM candidates.
We denote the set of observed paths as
\(I=\{1,\dots,L_{\mathrm{obs}}\}\) and CKM paths in $\mathbf{p}$ as \(J_{\mathbf{p}}=\{1,\dots,L(\mathbf{p})\}\).
We seek an injective mapping
\(\pi_{\mathbf{p}}: I\to J_{\mathbf{p}}\)
that assigns each observed path to a distinct CKM candidate.
The optimal assignment is obtained by
\begin{equation}
\pi_{\mathbf{p}}^\star \in
\arg\min_{\pi\in\mathrm{Inj}(I,J_{\mathbf{p}})}
\ \sum_{l\in I} D_{l,\pi(l)}(\mathbf{p}).
\label{eq:assign}
\end{equation}

Given the optimal assignment $\pi_{\mathbf{p}}^\star$, the matched distance for path $l$ with CKM in $\mathbf{p}$ is
\(D_{l,\pi_{\mathbf{p}}^\star(l)}(\mathbf{p})\) and the path weight
\(w_l(\mathbf{p}) = \frac{1}{1+D_{l,\pi_{\mathbf{p}}^\star(l)}(\mathbf{p})}\).
These weights quantify per-path consistency and are used by the subsequent geometric estimation.

To select the most confident match for each path globally in $\mathbf{p}\in\mathcal{G}_{\mathrm{cand}}$, we pick the best CKM grid $\mathbf{p}_l^\star$
\begin{equation}
\mathbf{p}_l^\star \in \arg\min_{\mathbf{p}\in\mathcal{G}_{\mathrm{cand}}} D_{l,\pi_{\mathbf{p}}^\star(l)}(\mathbf{p}),
\label{eq:best_overall_refined}
\end{equation}
and its paired path $m_l^\star=\pi_{\mathbf{p}_l^\star}^\star(l)$. Therefore, the global path weight of observed path $l$ is
\begin{equation}
w_l^* = \frac{1}{1+D_{l,\pi_{\mathbf{p}}^\star(l)}(\mathbf{p}_l^\star)}.
\label{eq:w_global}
\end{equation}

To make the angle–delay prior accessible for geometry, each CKM angle–delay pair can be geometrically mapped by (\ref{eq:ckm_scatterer_single_formula}) to a scatterer prior in the physical domain
\begin{equation}
(\theta_m^{\mathbf{p}},\ \tau_m^{\mathbf{p}})\ \mapsto\
\tilde{\mathbf{x}}_{s,m}^{\,\mathbf{p}}\in\mathbb{R}^2.
\label{eq:ckm_map}
\end{equation}

Collecting the global best matches $(\mathbf{p}_l^\star,\,m_l^\star)$ of observed path $l$ and the corresponding weights $w_l$, we form the selected CKM prior set
\begin{equation}
\mathcal{S}_{\mathrm{sel}}
=\Big\{\,\big(\tilde{\mathbf{x}}_{s,m_l^\star}^{\,\mathbf{p}_l^\star},\,w_l^*\big)\,\Big\}_{l=1}^{L_{\mathrm{obs}}}.
\label{eq:Ssel}
\end{equation}

Based on the distance in angle-delay power map, we further treat matches with credibility below a threshold as invalid. Specifically, candidates with $w_l < 0.5$ are regarded as mismatches and thus discarded from $\mathcal{S}_{\mathrm{sel}}$.

Intuitively, this step turns the strongest spectral peaks into high-confident scatterer locations, with credibility of $w_l^*$. The higher the $w_l$, the stronger influence in the subsequent solution.

\subsection{Joint Geometric Estimation with Priors}
\label{estimation}

From the previous subsection, each observed path now carries a CKM prior $\tilde{\mathbf{x}}_{s,m_l^\star}^{\,\mathbf{p}_l^\star}$ with its confidence weight $w_l^*$, together with the measured AoA and ToA in (\ref{eq:obs_set}). These form the input in this section. Our goal is to jointly estimate both the location of UE $\mathbf{x}_{\mathrm{UE}}$ and the corresponding scatterers $\{\hat{\mathbf{x}}_{s,l}\}$ under prior constraints. Given the BS location \(\mathbf{x}_{\mathrm{BS}}\),
each observed path~$l$ implies that its scatterer lies on the
BS-anchored ray
\begin{equation}
    \mathcal{L}_l=\{\ \mathbf{x}_{\mathrm{BS}}+d_l\,\mathbf{u}(\hat\theta_l)
\ \mid\ d_l\ge 0\ \},
\end{equation}
where \(\mathbf{u}(\hat\theta_l)=[\cos\hat\theta_l,\ \sin\hat\theta_l]^T\) is the unit direction vector.

We now estimate the UE location \(\mathbf{x}_{\mathrm{UE}}\)
together with ray-constrained scatterers
\(\{\hat{\mathbf{x}}_{s,l}\}\) by solving a weighted nonlinear
least-squares (NLS) problem.
The first term enforces consistency between geometry and the measured delays,
while the second term regularizes scatterers toward their CKM priors
with scatterer prior strength \(\lambda_{\mathrm{prior}}>0\)
\begin{equation}
\begin{aligned}
\min_{\mathbf{x}_{\mathrm{UE}},\,\{\hat{\mathbf{x}}_{s,l}\}}
&\sum_{l=1}^{L_{\mathrm{obs}}} w_l^*
\Big(\,\|\mathbf{x}_{\mathrm{UE}}-\hat{\mathbf{x}}_{s,l}\|_{2}
+\|\hat{\mathbf{x}}_{s,l}-\mathbf{x}_{\mathrm{BS}}\|_{2}
- c\,\hat\tau_l\,\Big)^2 \\
&\qquad\qquad
+\ \lambda_{\mathrm{prior}}
\sum_{l=1}^{L_{\mathrm{obs}}} w_l^*
\|\hat{\mathbf{x}}_{s,l}-\tilde{\mathbf{x}}_{s,m_l^\star}^{\,\mathbf{p}_l^\star}\|^2_{2} \\
\text{s.t.}\quad
&\hat{\mathbf{x}}_{s,l}\in\mathcal{L}_l,\qquad l=1,\dots,L_{\mathrm{obs}}.
\end{aligned}
\label{eq:nls_final_clean}
\end{equation}

This problem is solved iteratively using Levenberg--Marquardt scheme \cite{marquardt_algorithm_1963} with the UE initialized by the barycenter in~\eqref{eq:init_bary_global}, and each scatterer initialized at its CKM prior.

\section{Simulation Results}
The simulation results are provided in this section to quantitatively evaluate the localization performance of the proposed CKM-enabled method. The evaluation includes comparisons with an AoA–RSS fingerprinting baseline, sensitivity to receive array size, robustness to non-prior additional scatterers, and the resulting root mean square error (RMSE).

The proposed CKM-enabled localization method is evaluated under a two-dimensional NLoS environment, where multiple fixed scatterers are present and the direct LoS path is blocked. All simulations are conducted using the parameters summarized in TABLE~\ref{tab:sim_params}. The BS is placed at $(0,0)$. 15 fixed scatterers are randomly distributed within the rectangular region 
$[10,50]\times[-40,40]~\text{m}$, and UE test locations are randomly distributed within 
$[50,80]\times[-40,40]~\text{m}$.

The performance of the proposed method is compared against the following baselines:

\begin{itemize}
  \item \textbf{CKM Coarse Matching}: Employs CKM for coarse localization by matching the observed angle–delay signature with the CKM as~\eqref{eq:init_bary_global}.

  \item \textbf{AoA--RSS Fingerprint}~\cite{fingerprint}: Uses a conventional fingerprinting pipeline that matches the observed AoA/RSS feature vector to an offline database.
\end{itemize}

\begin{table}[t]
\vspace*{0.08in} 
\centering
\caption{Simulation Parameters}
\label{tab:sim_params}
\begin{tabular}{cc}
\toprule
\textbf{Parameter} & \textbf{Setting} \\
\midrule
Carrier frequency  & 6 GHz \\
Bandwidth & 100 MHz \\
Number of subcarriers  & 1024 \\
FFT size & 1024 \\
SNR Value & 30 dB \\
Reference point interval of CKM & 1 m \\
Scatterer prior strength $\lambda_{\mathrm{prior}}$ & 2 \\
Number of candidate CKM $K_{\mathrm{cand}}$ & 10 \\
DFT size $N_\theta$ in angle domain & 256 \\
DFT size $N_\tau$ in delay domain & 1024 \\
\bottomrule
\end{tabular}
\end{table}

\begin{figure}[t]
    \centering
    \includegraphics[width=0.7\linewidth]{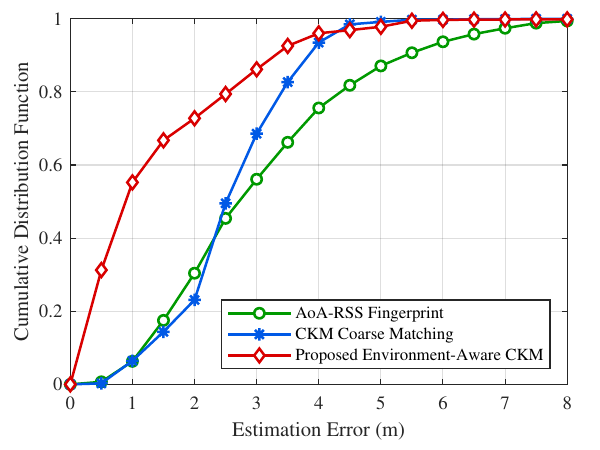}
    \caption{CDF of localization error with different method when $M=32$.}
    \label{fig:CDF_3types}
\end{figure}

\begin{figure}[t]
    \centering
    \includegraphics[width=0.7\linewidth]{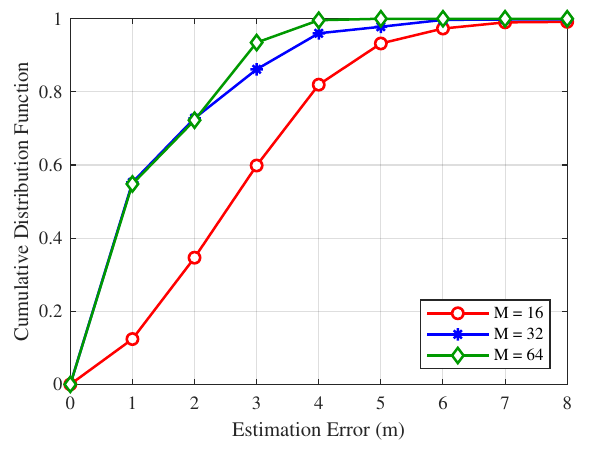}
    \caption{CDF of localization error with different receive antennas $M$.}
    \label{fig:Different_Nrx}
\end{figure}

\begin{figure}[t]
    \centering
    \includegraphics[width=0.7\linewidth]{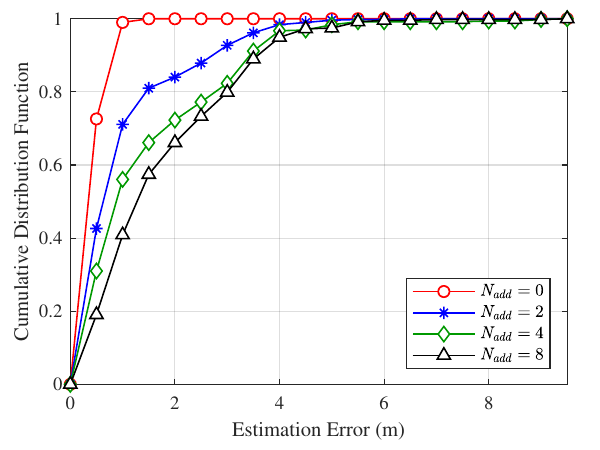}
    \caption{CDF of localization error with additional non-prior scatterers $N_{add}$ when $M=32$.}
    \label{fig:Different_nnew}
\end{figure}


Fig.~\ref{fig:CDF_3types} shows the cumulative distribution function (CDF) of localization error for different methods when \(M=32\). To further verify the robustness of the proposed framework, we additionally placed four non-prior scatterers in the environment as interference. It can be observed that the curve of the proposed environment-aware CKM method almost lies in the upper-left of the two baseline methods, representing higher overall accuracy. At the error level of \(2.5\,\mathrm{m}\), it already reaches nearly 80\% reliability. The advantage lies in the fact that the proposed method not only utilizes the channel knowledge but also exploits the geometric prior of environmental scatterers in the CKM, thereby resolving the ambiguity in coarse localization and achieving more refined estimation. In addition, CKM coarse matching is overall superior to the  AoA--RSS fingerprint, because the former integrates both AoA and ToA information which have higher resolution compared with AoA and RSS.

Fig.~\ref{fig:Different_Nrx} also presents the error CDF under different numbers of receive antennas \(M\). It can be observed that as \(M\) increases, localization accuracy consistently improves, which reflects higher angular resolution and better multipath separation. In the region where the error is less than \(2\,\mathrm{m}\), the curves for \(M=64\) and \(M=32\) almost overlap, indicating that in this scenario, \(M=32\) is sufficient to resolve the dominant multipath. When the error is greater than \(2\,\mathrm{m}\), \(M=64\) performs better due to its stronger resolving capability.

\begin{figure}[t]
    \centering
    \includegraphics[width=0.75\linewidth]{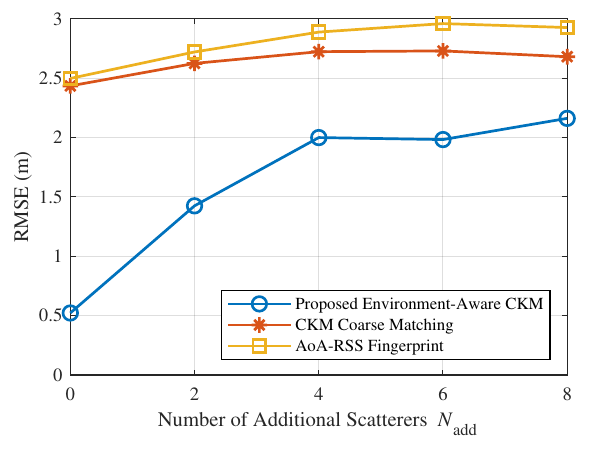}
    \caption{Localization accuracy comparison with different method for additional non-prior scatterers $N_{\text{add}}$ when $M=32$.}
    \label{fig:Different_nnew_RMSE}
\end{figure}

To simulate a time-varying scattering environment, we evaluated the performance under different numbers of additional non-prior scatterers \(N_{\text{add}}\). Fig.~\ref{fig:Different_nnew} shows that as \(N_{\text{add}}\) increases, the CDF curve of the proposed method shifts to the right, which means the performance decreases. However, when \(N_{\text{add}}=0\), the reliability within \(1\,\mathrm{m}\) can reach about 99\%. Even when \(N_{\text{add}}=8\), the reliability within \(2\,\mathrm{m}\) still approaches 65\%. Fig.~\ref{fig:Different_nnew_RMSE} provides the RMSE comparison of different methods. When \(N_{\text{add}}=0\), the RMSE of the proposed method is about \(0.5\,\mathrm{m}\), while the AoA--RSS fingerprint and CKM coarse matching are around \(2.5\,\mathrm{m}\). When \(N_{\text{add}}\) increases to 8, the RMSE of the proposed method rises to about \(2.2\,\mathrm{m}\), still significantly better than the two baselines. Fig.~\ref{fig:Different_nnew} and Fig.~\ref{fig:Different_nnew_RMSE} together demonstrate that introducing the geometric prior of the environment in the environment-aware CKM is crucial for improving localization accuracy and robustness under time-varying scenarios.

\vspace{-2mm}
\section{Conclusion}
\vspace{-1mm}
In this paper, we propose an environment-aware CKM localization framework for complex NLoS environments. The method leverages AoA--ToA information in the CKM and the geometric prior of environmental scatterers, effectively eliminating ambiguities in the coarse localization stage and achieving more refined localization. Unlike fingerprint-based approaches, the proposed scheme can reliably distinguish multipath and maintains strong robustness in the presence of additional unknown scatterers. Simulation results show that the proposed method significantly outperforms the CKM coarse matching and AoA--RSS fingerprint baselines in different scenarios. Under interference conditions, the framework still achieves sub-meter localization accuracy, fully demonstrating its effectiveness and robustness in practical localization applications.
\vspace{-1mm}
\section*{Acknowledgment}
\vspace{-1mm}
This work was supported by the National Science and Technology Major Project under Grant 2025ZD1304500, and by the National Natural Science Foundation of China under Grant 62571116 and 62171127, and by the Fundamental Research Funds for the Central Universities under Grants 2242022k60004 and 3204002004A2.
\vspace{-1mm}

%



\ifCLASSOPTIONcaptionsoff
  \newpage
\fi



\bibliographystyle{IEEEtran}
%

\bibliography{test}

%




\end{document}